\documentclass{sig-alternate}

\def\linksnum{38,282}
\def\linkspp{3,553}
\def\linksppper{9.28}
\def\linkscp{34,552}
\def\linkscpper{90.26}
\def\linksss{177}

\def\paths{733,931}
\def\nodes{18,943}

\newenvironment{description*}
  {\begin{description}
    \setlength{\itemsep}{0pt}
    \setlength{\parskip}{0pt}}
  {\end{description}}

\newenvironment{enumerate*}
  {\begin{enumerate}
    \setlength{\itemsep}{0pt}
    \setlength{\parskip}{0pt}}
  {\end{enumerate}}

\def\pathsred{12.49}

\def\linksnumred{4.34}
\def\nodesred{1.87}

\def\pathsav{838,734}
\def\linksav{40,023}
\def\nodesav{19,305}

\def\assnum{38}
\def\seentruenum{3,724}

\def\seeninfpp{623}
\def\seeninfss{31}
\def\seeninfcppc{3,070}

\def\seentrueppper{16.1\%}
\def\seentruessper{1.2\%}
\def\seentruecppcper{82.6\%}

\def\validnumpp{516}
\def\validnumss{28}
\def\validnumcppc{2,964}

\def\validnumtot{3,508}

\def\validinfpertot{94.2\%}

\def\validinfperpp{82.8\%}
\def\validinfperss{90.3\%}
\def\validinfpercppc{96.5\%}

\def\asswfulldata{27}
\def\truelinks{1,114}
\def\seenlinks{552}

\def\truepplinks{865}

\def\trueseenppdiffper{38.7\%}

\def\truesslinks{30}

\def\trueseenssdiffper{93.3\%}
\def\truecppclinks{218}

\def\trueseencppcdiffper{86.7\%}
\def\ppper{77.6\%}

\def\ssper{2.7\%}
\def\cppcper{19.6\%}

\usepackage{rotating,url,graphicx,epsfig,subfigure,amsmath,multirow}

\begin{document}
\title{AS Relationships: Inference and Validation}

\author{
Xenofontas Dimitropoulos\\ Georgia Tech/CAIDA \\ fontas@ece.gatech.edu
\and Dmitri Krioukov \\ CAIDA \\ dima@caida.org
\and Marina Fomenkov \\ CAIDA \\ marina@caida.org
\and Bradley Huffaker \\ CAIDA \\ brad@caida.org
\and Young Hyun \\ CAIDA \\ young@caida.org
\and kc claffy \\ CAIDA \\ kc@caida.org
\and George Riley \\ Georgia Tech \\ riley@ece.gatech.edu}

\maketitle

\begin{abstract}

Research on performance, robustness, and evolution of the global
Internet is fundamentally handicapped without accurate and thorough
knowledge of the nature and structure of the contractual relationships
between Autonomous Systems~(ASs). In this work we introduce novel
heuristics for inferring AS relationships. Our heuristics improve
upon previous works in several technical aspects, which we outline
in detail and demonstrate with several examples.
Seeking to increase the value and reliability of our
inference results, we then focus on validation of inferred
AS relationships.
We perform a survey with ASs' network administrators to
collect information on the actual connectivity and policies of the
surveyed ASs. Based on the survey results, we find that our new AS
relationship inference techniques achieve high levels of
accuracy: we correctly infer \validinfpercppc~customer to
provider~(c2p), \validinfperpp~peer to peer~(p2p), and
\validinfperss~sibling to sibling~(s2s) relationships. We then
cross-compare the reported AS connectivity with the AS connectivity
data contained in BGP tables. We find that BGP tables
miss up to~86.2\% of the true adjacencies of the surveyed ASs. The
majority of the missing links are of the p2p type, which highlights
the limitations of present measuring techniques to capture links of
this type. Finally, to make our results easily accessible
and practically useful for the community, we open an AS relationship
repository where we archive, on a weekly basis, and make publicly
available the complete Internet AS-level topology annotated with AS
relationship information for every pair of AS neighbors.

\end{abstract}


\section{Introduction}

The global Internet routing system is composed of thousands of Autonomous Systems
(ASs) that operate individual parts of the Internet infrastructure. ASs engage in
a variety of relationships to collectively and
ubiquitously route traffic in the Internet. These relationships are usually
realized in the form of business agreements that, in turn, translate into
engineering constraints on traffic flows within and across individual networks.

Understanding the underlying business AS relationships plays a critical role in
many research and operational tasks ranging from realistic simulations of
packets routed in the Internet to selection of peers or upstream providers based
on connectivity and AS relationships of candidate ISPs. Further, statistical data
on these relationships are useful for development of more advanced interdomain
routing protocols and architectures that take into account the presence of
AS relationships to improve their performance~\cite{SuKaEe05}. Moreover,
business behavior patterns of Internet players influence directions of ISPs'
collaboration and ultimately the evolution of the macroscopic infrastructure
of the Internet.

In this study we follow previous works~\cite{Gao01,SuAgReKa02,DiBaPaPi03,ErHaSch02}
in considering the following three major categories of AS relationships:
{\em customer-to-provider} (c2p), {\em peer-to-peer} (p2p), and
{\em sibling-to-sibling} (s2s). In the c2p category, a customer AS pays a
provider AS for any traffic sent between the two.\footnote{
We use acronym {\em c2p\/} to refer to customer to provider relationships in
general, as well as to links $A$-$B$, where AS~$A$ is a customer of AS~$B$. In
contrast, we use acronym {\em p2c\/} to refer only to links $A$-$B$, where
AS~$A$ is a provider of AS~$B$.} In the p2p category, two ASs
freely exchange traffic between themselves and their customers, but do not
exchange traffic from or to their providers or other peers. In the s2s category,
two ASs administratively belong to the same organization and freely exchange
traffic between their providers, customers, peers, or other siblings.

Our work makes the following contributions:
\begin{enumerate*}

\item We introduce novel heuristics for inferring c2p, p2p, and s2s relationships.
Our heuristics improve the state-of-the-art in several technical aspects, one of
them being a more realistic problem formulation that accepts that AS paths
do not always exhibit a hierarchical pattern. We demonstrate using several
examples our enhancements that lead to more accurate inference results.

\item We conduct a survey with organizations operating ASs, from which we
retrieve company-verified information about the actual types of relationships
they have with other networks. We use this information to validate the AS
relationships we infer and find that they are highly accurate. To our
knowledge, this study is the most exhaustive AS relationship validation
effort to date.

\item Using company-verified data we confirm previous measurement
results~\cite{ChaGoJaSheWi04,RaCo06} on the poor coverage of AS
topologies. In addition, we verify the commonly held assumption that
most of the missing links are of p2p type.

\item To promote further analysis and discussion of the macroscopic Internet
topology, we introduce a publicly available AS relationships
repository~\cite{as-rel-data}. We automate our heuristics and archive datasets of
annotated AS links on a weekly basis. We also compute and publish ranking of
ASs based on inferred AS relationship hierarchies~\cite{as-ranking}.

\end{enumerate*}

This paper follows our earlier work~\cite{DiKrHuClRi05} on inferring
c2p relationships. It addresses the issue left open of how to select
the most realistic from the candidate solutions to our c2p problem
formulation. It then extends our previous work by: 1)~introducing new
heuristics for the inference of p2p and s2s relationships, 2)~validating
our inferences, and 3)~developing an open AS relationships repository.

We organize the paper as follows.  In the next section we introduce
and describe in detail our heuristics. We compare our approach to
inferring AS relationships with previous ones and discuss our
improvements.  In section~\ref{sec:results} we
apply the developed heuristics to Internet data and fully annotate a
snapshot of the AS topology with the computed types of
relationships. We also briefly discuss our ranking of ASs based on
inferred AS relationship hierarchies. In section~\ref{sec:sur}, we
describe the results of our AS survey, validate our heuristics, and
analyze the true AS relationships that we learned from the
participating ASs. Finally, we conclude in section~\ref{sec:con}.

\section{Inference Heuristics}\label{sec:algos}
\subsection{Preliminaries}
\label{sec:algos:pre}

Gao's seminal work~\cite{Gao01} was the first to formulate and systematically
study the AS relationships inference problem. Gao assumed that every BGP path
must comply with the following hierarchical pattern: an uphill segment of zero
or more c2p or s2s links, followed by zero or one p2p links, followed by a
downhill segment of zero or more p2c or s2s links. Paths with this
hierarchical structure are {\em valley-free} or {\em valid}. Paths that
do not follow this hierarchical structure are called {\em invalid} and
may result from BGP misconfigurations or from BGP policies that are more
complex and do not distinctly fall into the c2p/p2p/s2s classification.
Following this
definition of valid paths, Gao proposed an inference heuristic (which we denote
as GAO) that identified top providers and peering links based on AS degrees and
valid paths.

Following Gao's work, Subramanian {\it et al}.~\cite{SuAgReKa02}
developed a mathematical formulation of the inference
problem. They cast the inference of AS relationships into the
{\em Type of Relationship}~(ToR) combinatorial optimization problem: given a
graph~$G(V,E)$ derived from a set of BGP paths~$P$, assign the edge
type (c2p or p2p; s2s relationships are ignored) to every edge \mbox{$i \in E$}
such that
the total number of valid paths in~$P$ is maximized. The authors speculated that
ToR is NP-complete and developed a heuristic solution, which we refer to as SARK below.

The SARK approach takes as input the BGP tables collected at different vantage
points and computes a {\it rank} for every AS. This rank is a measure of how
close to the graph core an AS lies and it is equivalent to node
{\em coreness}~\cite{AlDaBaVe04,AlDaBaVe05}. The heuristic then infers AS relationships
by comparing ranks of adjacent ASs. If the ranks are similar, the algorithm
classifies the link as p2p, otherwise it is c2p.

Di~Battista {\it et al}.~\cite{DiBaPaPi03} and Erlebach
{\it et al}.~\cite{ErHaSch02} independently showed that ToR is
indeed NP-complete, developed mathematically rigorous approximate
solutions to the problem and proved that it is impossible to infer p2p
relationships under the ToR formulation framework. For this reason, their
solutions (referred to as DPP and EHS) infer c2p relationships
only and ignore p2p and s2s relationships.

Despite the ToR formulation being substantially studied, we find that
it bears limitations that lead to incorrect inferences. We describe
these limitations with the following examples.

\begin{figure}
    \centering
    \includegraphics[width=2.3in]{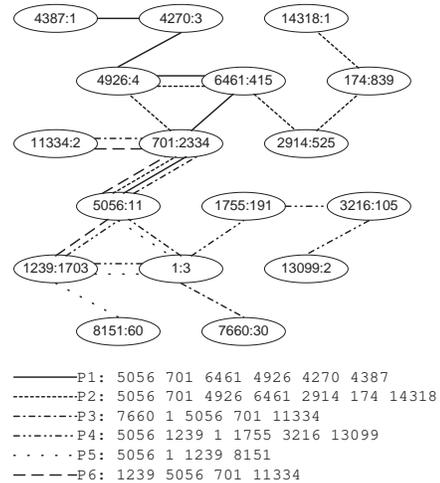}
    \caption{\scriptsize An instance of the ToR problem that does not admit a solution.
    Each circle is marked with a tuple~$X:Y$, where~$X$ is the AS number and~$Y$ is the
    AS degree seen in our AS topology. The paths at the bottom yield the ToR instance.}
    \label{fig:koshmar}
\end{figure}

{\bf \emph{Example 1.} Ignoring s2s relationships causes proliferation of erroneous inferences.}
Consider a path~$p=\{ij\} \in P$ that includes
an edge~$i$ appearing in multiple paths, and an edge~$j$, appearing only in this
path~$p$. Suppose that in reality~$i$ is a sibling edge and~$j$ is a c2p edge.
It is convenient to represent a c2p edge annotation by making the edge directed
{\em from\/} the customer AS {\em to\/} the provider AS. Depending on the structure
of other paths containing the sibling edge~$i$, a ToR solution can direct it as either
c2p or p2c. If it gets directed as p2c, then to make the path~$p$ valid, the algorithm
has to erroneously direct the edge~$j$ as p2c, too.

{\bf \emph{Example 2.} A solution maximizing the number of inferred-to-be-valid
paths is not necessarily correct.}
Consider the real-world instance of the ToR problem in Figure~\ref{fig:koshmar},
which was used in~\cite{DiBaPaPi03} to introduce ToR.
In this setting there are four distinct combinations of edge orientations,
each maximizing the number of valid paths, but rendering one of the paths P1,
P2, P4, or P5 as invalid. In path P6, AS 5056 with degree~11 appears to transit
traffic between large providers, AS 701 (UUNET) and AS 1239 (Sprint). A ToR
solution will treat path P6 as valid. Thus, it must infer AS 5056
as a provider of either UUNET or Sprint, or a provider of both,
all of which are incorrect. The key point of this example is that while
it is reasonable to assume that most AS paths in the Internet have a valid
hierarchical structure, it is still possible that some paths
in real networks have a non-hierarchical invalid structure. An attempt to annotate such paths based on
the valley-free model will result in unrealistic relationships.

{\bf \emph{Example 3.} In cases when there are multiple solutions with
the same number of valid paths, ToR has no means to deterministically
select the most realistic solution.} Instead, it has to randomly attribute
validity to one of the available solutions.
Consider path~\mbox{$p \in P$} that is a sequence of edges
\mbox{$i_1,i_2,\ldots,i_{|p|-1},j \in E$}. Suppose that the last edge~$j$
appears only in this one path~$p$ and that it is from a large provider (such
as UUNET) to a small customer. Suppose that other edges
\mbox{$i_1,i_2,\ldots,i_{|p|-1}$} appear in several other paths and are
correctly inferred as c2p. In this scenario both orientations of the edge~$j$
(i.e.~correct p2c and incorrect c2p) render path~$p$ valid. Thus, this edge
cannot receive a deterministic direction from the ToR solution. This example
explains why Rimondini~\cite{rimondini02} found several well-known large
providers such as AT\&T, Sprint, Level3 to be inferred as customers of
smaller ASs such as AS2685~(degree 2), AS8043~(1) and AS13649~(7),
respectively. We also observed incorrect inferences of this
type in our experiments.

The above examples illustrate that:
1)~it is necessary to account for s2s relationships;
2)~trying to simply maximize the number of valid AS paths may result in
incorrect AS relationship inferences; and
3)~without additional information, the ToR framework by itself is insufficient
to ensure a deterministic inference of AS relationships.

In the following subsections we address these shortcomings and present
heuristics to determine AS relationships more accurately.

\subsection{Inferring s2s relationships}
\label{sec:algos:s2s}

Sibling links connect ASs belonging to the same organization, and thus communication
between s2s ASs is not subject to the export restrictions found in c2p and p2p
relationships. For example, rules such as ``prefixes learned from a peer cannot be
announced to other peers'' do not apply for sibling ASs. Therefore, sibling ASs
can implement much more flexible and diverse policies than non-affiliated ASs,
making it very difficult to infer s2s relationships from BGP data. For this
reason we utilize the IRR databases to annotate s2s links. We then remove s2s
edges from both graph~$G$ and path set~$P$ to avoid proliferation of incorrect
c2p inferences. In effect, we abrogate the limitations of Example~1 by independently
inferring s2s relationships.

Specifically, we track the organization to which each AS is registered in the
databases and create groups of sibling ASs registered to the same organization.
In several cases sibling ASs are registered to syntactically different
organizational names, which still represent the same organization by other
measures.  For example, ASs 7018 and 3339 are registered to ``AT\&T WorldNet
Services'' and ``AT\&T Israel'', respectively. To find such cases, we examine
the organization names and manually create a dictionary of organization name synonyms.
Then, we infer as s2s the ASs that are registered to the same organization name
or to synonymous organization names.

The strength of our approach is that it takes advantage of explicit information
contained in the IRR databases. Although we realize these databases are not
always up-to-date or perfectly accurate, the organization names change less
frequently than BGP policies and other more dynamic attributes. We can therefore
treat the IRR databases as a source of publicly available information, which is
reasonably accurate for the purpose of inference of s2s relationships.

\subsection{Improving the integrity of c2p inferences}
\label{sec:algos:c2p}

In Example~2 we demonstrated that trying to maximize the number of
inferred-to-be-valid paths can lead to incorrect inferences since
in reality AS paths are not always hierarchical. To address this
limitation we construct a c2p-inference heuristic that is based on
the idea of relaxing the requirement for a maximal number of valid paths
and using the AS degree information to detect paths that are invalid
and that we should not try to direct as valid. We formalize this idea as
follows.

For every edge \mbox{$i \in E$} we introduce a
weight~$f(d_i^-,d_i^+)$ that is a function of the
degrees~$d_i^-$ and~$d_i^+$ (\mbox{$d_i^- \leq d_i^+$})
of the ASs adjacent to the edge~$i$. The weight $f$ is large when there
is a significant degree difference ($d_i^- \ll d_i^+$)
between these neighboring ASs, and small otherwise. In
directing the edges of the graph, we use~$f$ in the
following way: when an edge $i$ is directed
from a small-degree AS to a large-degree AS,
it earns a {\em bonus} $b_i$ equal to~$f(d_i^-,d_i^+)$,
otherwise $b_i = 0$.
We then formulate the inference problem as the following
multiobjective optimization problem:

\begin{description*}
\item[$O_1$] Maximize the number of valid paths in~$P$;
\item[$O_2$] Maximize the sum $\sum_{i \in E} b_i$.
\end{description*}

These two methodological objectives can be conflicting. Consider again the
Example~3 using Figure~\ref{fig:koshmar}. According to the objective~$O_1$, at least
one of the edges 1239-5056 or 5056-701 in~P6 must be directed against the node
degree gradient in order to render P6 valid. By introducing the second
objective~$O_2$, we relax the first objective's requirement for the maximal
number of valid paths. We can thus accept an ``invalid'' orientation for~P6
based on the strong degree-gradient indication~($O_2$) that neither 1239 nor
701 are customers of~5056.

This formulation combines the strengths of previous works. First, it is similar
to SARK, DPP and EHS, in that it respects the valley-free model and tries to
maximize the number of valid paths in the input path set~$P$. Secondly, it
is similar to GAO, in that it uses the implicit knowledge embedded in AS degree
information to assign directions to edges along the node degree gradient
by giving certain weighted preference to edge orientations collinear with
this gradient.

To solve the newly formulated optimization problem, we map the c2p or p2c
relationship of edge~$i$ to boolean variable~$x_i$ as follows:
assuming an arbitrary initial direction of~$i$, an assignment of {\em true\/} to~$x_i$
means that edge~$i$ keeps its original direction, while an assignment of {\em
false\/} to~$x_i$ reverses the direction of~$i$. We find assignments
to variables~$x_i$ by reducing the multiobjective optimization
problem to the well-known MAX2SAT problem.

MAX2SAT is a boolean algebra problem: given a set of clauses with two boolean
variables per clause~\mbox{$l_i \vee l_j$}, find an assignment of values to
variables maximizing the number of simultaneously satisfied clauses~\cite{GaJo79}.
If the clauses are weighted, the problem is to maximize the sum of weights of
the simultaneously satisfied clauses. MAX2SAT is NP-complete, however, the
semidefinite programming~(SDP) approach~\cite{GoWi95} delivers an approximate
answer that differs from the exact answer by not more than a factor of~0.94.

To reduce the objective~$O_1$ (ToR) to MAX2SAT we use the approach of DPP and
EHS~\cite{DiBaPaPi03,ErHaSch02}. This gives a set of~$x_i \vee x_j$ clauses, where
$i,j \in E$.

To reduce the objective~$O_2$ to MAX2SAT, we introduce a clause
\mbox{$x_i \vee x_i$} for every edge~$i \in E$ that has an initial direction
along the node degree gradient, and a clause \mbox{$\bar{x_i} \vee \bar{x_i}$}
for every edge with an initial direction against the node degree gradient. We
thus ensure that if an edge is directed along the node degree gradient, then
the corresponding clause is satisfied. To make our MAX2SAT instance equivalent
to~$O_2$, we weight every clause by~$b_i=f(d_i^-,d_i^+)$.

We then reduce the resulting multiobjective optimization problem to MAX2SAT by
refining the weights of the clauses. We introduce a parameter~$\alpha$ and
weight the objective~$O_1$ by~$\alpha$ and the objective~$O_2$ by~$1 - \alpha$:
\begin{equation}\label{eq:weights:2}
    w_{ij}(\alpha) =
    \begin{cases}
        c_1\alpha                   & \text{for $O_1$ clauses,}\\
        c_2(1-\alpha)f(d_i^+,d_i^-) & \text{for $O_2$ clauses.}
    \end{cases}
\end{equation}
The normalization coefficient~$c_1$ is determined from the condition
$\sum_{i \neq j} w_{ij}(\alpha)=\alpha$ $\Rightarrow$ $c_1 = 1/m_1$,
where~$m_1$ is the number of $O_1$~clauses.
The normalization coefficient~$c_2$ is determined from the condition
$\sum_iw_{ii}(\alpha)=1-\alpha$.
Varying~$\alpha$ in the region between~0 and~1 controls the relative
preference of the two objectives.\footnote{In the terminology of multiobjective
optimization~\cite{CoSi03}, we consider the simplest scalar method of weighted sums.}
We explore the tradeoff between the objectives~$O_1$ and~$O_2$ and
adjust~$\alpha$ to the region or the point that results in the most accurate
AS relationship inferences (cf. discussion of the optimal value of $\alpha$ in
section~\ref{sec:results:relat}).

Function~$f$ encodes dependence on AS degrees into our inference process. This
function should take large values when its two degree arguments differ
significantly, otherwise its values should be small, because
neighboring ASs with significant size difference typically have a customer to
provider relationship and AS size is strongly correlated to AS degree~\cite{TaDoGoJaWiSh01}.
We note that a given
absolute difference in AS degrees is of different importance for small ASs and
for large ASs. For example, a degree difference of~50 says more about the
relative size of two ASs of degrees~1 and~51, than of~3000 and~3050. To account
for this relative importance, we normalize the degree difference in~$f$ to the
{\em relative\/} node degree gradient~\mbox{$(d_i^+-d_i^-)/(d_i^++d_i^-)$}.
In addition, topology graphs derived from BGP data provide only approximations
of the true AS degrees. They tend to underestimate degrees of small ASs but
yield more accurate degree approximations for larger ASs~\cite{ChaGoJaSheWi04}.
To model this effect, we introduce a logarithmic factor reflecting our stronger
confidence in accuracy of large AS degrees, compared to small ones. We thus
construct~$f$ as:
\begin{equation}
    f(d_i^+,d_i^-)=\frac{d_i^+-d_i^-}{d_i^++d_i^-}\log(d_i^++d_i^-).
\end{equation}

In summary, our formulation of the c2p relationship inference problem exploits
the structure of the AS paths to address the limitations that we illustrated in
Examples~2 and~3 of section~\ref{sec:algos:pre}.

\subsection{Inferring p2p relationships}
\label{sec:algos:p2p}

The inference of p2p relationships is more challenging than the inference of
c2p relationships. As both DPP and EHS show, it is impossible to infer p2p
relationships within the ToR formulation framework. Indeed, a valid path can
have only one p2p link adjacent to the top provider in the path. If we replace
this p2p link with a c2p or p2c link, the path remains valid, as it still has
a valley-free, hierarchical structure. Therefore, maximizing the number of
valid paths as is done by ToR, one cannot deterministically infer any
p2p relationships at all. Confirming the difficulty of inferring p2p relationships
comes a work by  Xia and Gao~\cite{XiaGao04}, who find that GAO and SARK's p2p
inference heuristics yield a low accuracy of, respectively, 49.08\% and 24.63\%
of correct p2p inferences.

To improve the inference of p2p relationships, we develop a
heuristic that combines GAO and DPP strengths. We start from a set of BGP
paths~$P$ and extract a graph~$G$ from it. Then we preprocess~$P$ to identify
links that are not of p2p type~(non-p2p).

According to the valley-free model, a path can have at most one p2p link and
this link must be adjacent to the top provider of the path. We thus parse all
paths in~$P$ and denote all links that are not adjacent to the highest degree AS
in a path as non-p2p.
This approach is similar but not identical to GAO. GAO assumed that 1) a p2p
link can lie only between the highest degree AS in a path and its highest degree
neighbor and 2) that the degree ratio between the two edge ASs of a p2p link is
smaller than an external parameter (discussed below).
This method is aggressive in excluding non-p2p links. To illustrate,
consider an AS path A-B-C-D, where AS degrees are~$d_A = 10$, $d_B = 500$,
$d_C = 1000$, and~$d_D = 501$. GAO allows only link C-D to be of p2p type and
denotes the others as non-p2p. However, the degree difference between B and D is
too small to make this judgment reliably. Our heuristic addresses this
shortcoming by including both B-C and C-D as candidate p2p links. We denote
by~$R$ the set of possible p2p edges constructed this way.

We then introduce a weight~$g(d^-_i,d^+_i)$ for every edge~$i \in R$. Weight~$g$
is large when the ASs adjacent to the edge~$i$ have similar degrees, and small
otherwise. Such weighting expresses our higher confidence that a pair of
neighboring ASs are peers when their degrees are similar. Our selected weight~$g$
complements the weight~$f$ used for the inference of c2p links:
\begin{equation}
    g(d_i^-,d_i^+)= 1 - c_3f(d_i^-,d_i^+),
\end{equation}
where $c_3=1/\max_{i \in E} f(d_i^-,d_i^+)$ is a normalization
coefficient.

Next, we remove from~$R$ any links that connect ASs with large degree
differences~$d^-_i \ll d^+_i$. More specifically, we introduce a threshold~$w_e
\in [0,1]$ and remove every edge~$i$ with~$g(d_i^-,d_i^+) < w_e$. The GAO
heuristic used an empirically selected value of~60 or~$\infty$ for a similar
threshold. We improve upon this approach by using information learned from
our survey (see section~\ref{sec:sur}) to choose a proper value for~$w_e$.
Namely, for each true p2p and c2p link present both in our survey results and
in~$R$, we examine what selection of~$w_e$ leads to:
1)~erroneously excluding a true p2p link from the set of possible p2p links~$R$,
meaning that $g(d^-_i,d^+_i) < w_e$ for a true p2p link;
and
2)~erroneously not excluding a true c2p link from the set of possible p2p
links~$R$, meaning that $g(d^-_i,d^+_i) > w_e$ for a true c2p link.
We find that the value of~$w_e$ that minimizes errors is~$g(3,545)$. The need
for external threshold~$w_e$ is unfortunate, but the large degree difference
between~$d^-=3$ and~$d^+=545$ indicates that this threshold simply cleans~$R$
of links that are unlikely to be of p2p type.

At the last step of our p2p inference process, we examine those paths in~$P$
that contain more than one edge from~$R$. Such paths violate the valley-free
model, and we need to classify some links from~$R$ as non-p2p in order to
resolve this violation. DPP showed that the problem of finding a maximal set
of p2p links that do not introduce invalid paths in~$P$ is equivalent to the
Maximum Independent Set~(MIS) problem. In the MIS formulation, we are given a
graph with nodes in~$N$ and arcs in~$A$ and we need to find the maximum subset
of~$N$ such that no two nodes of the subset are joined by an arc in~$A$. To
increase the reliability of the p2p link determination, we utilize our assigned
link weights~$g$ and turn the MIS problem into the Maximum Weight Independent
Set~(MWIS) problem. In the MWIS formulation, we give preference to edges with
large weights because we know that these edges are more likely to be of p2p
type. We solve the NP-complete MWIS problem by means of a polynomial time
approximation~\cite{qualex} and find a maximal weight subset of~$R$ that does
not create invalid paths in~$P$. We denote this subset as~$F$ and admit it as
our final set of p2p links.

\subsection{Summary of inference heuristics}
\label{sec:algos:sum}

In summary, our inference heuristics take as input a set of BGP paths~$P$ and a
corresponding graph~$G(V,E)$ and perform the following three consecutive
steps:
\begin{enumerate*}
\item Use IRRs to infer s2s relationships and create set~\mbox{$S \subset E$}
of s2s links;
\item Remove the subset $S$ from consideration and apply our heuristic
assigning c2p/p2c relationships to the links remaining in~$E \setminus S$;
\item Use~$P$ and~$G$ to infer p2p relationships and to create set~$F \subset
E$ of p2p links.
\end{enumerate*}
The final result is set~$S$ of s2s links, set~$F$ of p2p links, and set~$E
\setminus F \setminus S$ of c2p links.

\subsection{Related work}
\label{sec:algos:rel}

In comparison with other approaches to AS relationship inference, our heuristics
offer a number of improvements. In contrast to DPP~\cite{DiBaPaPi03} and
EHS~\cite{ErHaSch02}, we identify not only c2p, but p2p and s2s relationships
as well. Moreover, our c2p heuristic addresses the limitations we discussed
in section~\ref{sec:algos:pre} with ToR solutions.

The work by Subramanian {\it et al}.~\cite{SuAgReKa02}
introduced the ToR problem and the SARK
heuristic~\cite{SuAgReKa02} for solving ToR. SARK used node
coreness~\cite{AlDaBaVe04,AlDaBaVe05}, which reflects ASs' topological
positions in AS graphs, as a metric for inferring c2p and p2p
relationships. In contrast, our heuristics use AS degrees and policies
encoded in AS paths to infer c2p and p2p relationships.

The work by Gao~\cite{Gao01} used AS degrees and the
valley-free model to infer c2p, p2p, and s2s relationships. GAO algorithm treats every AS path as a hint of true
types of links in the path. It takes a set of AS paths as input, directs every link
toward the highest degree AS in the path, and after parsing all paths, counts the
directions each link has accumulated. If a link has received consistent directions
throughout the process, it is marked as c2p with the provider being at the top of
the directed link. Otherwise, the link is marked as s2s. Similarly to this work,
our heuristics employ the valley-free model and AS degrees to infer c2p and p2p
relationships, but we make a number of
technical enhancements, which we outline in detail in sections~\ref{sec:algos:c2p}
and~\ref{sec:algos:p2p}.
We use the IRR databases to infer s2s relationships, since it is hard
to reliably infer them from BGP data.

Xia and Gao~\cite{XiaGao04} used the IRR databases to extract
relationships among a subset of ASs and proposed a variation of the GAO heuristic
that takes this subset as an input to infer other AS relationships. They
demonstrated how accurate and current IRR databases provide explicit information
on AS relationships. On the other hand, dealing with IRR data has its own
intrinsic methodological problems: 1)~it is much harder to automate;
2)~the data is not always accurate and its accuracy level is hard
to estimate; 3)~not all ASs are registered.
In our work we also use the IRR data but only for s2s relationship inference.
For this task, we process the organization description records, which
are relatively stable over time, compared to policy-related records.

Mao {\it et al.}~\cite{MaQiEaZh05} proposed an AS relationship
inference technique that employs the valley-free model to infer c2p
and p2p relationships.
The technique introduces a set of new interesting ideas based on
the assumption that ASs prefer shorter AS paths over longer AS paths. This
assumption does not however hold when ASs use routing policies to select the
next-hop AS on the basis of its policy ranking, regardless of AS path
lengths.

Recent work by Muehlbauer {\it et al.}~\cite{MuFeMaRoUh06} introduced a shift from
inferring AS relationships to inferring AS paths using a model with {\em agnostic\/}
AS relationships and multiple routers per AS. They found that their model leads to
more accurate results, as far as accuracy of capturing path diversity is concerned,
than a model using inferred AS relationships and a single router per AS. By definition,
agnostic approaches cannot however capture precise characteristics of individual ASs.
Therefore, agnostic approaches are not appropriate for tasks such as constructing realistic
economy-based evolution models of ASs.
In addition, \cite{MuFeMaRoUh06,UhMaMu06} assumed
that models with c2p/p2p/s2s relationships are equivalent to models with a
single router per AS. The former models can however be extended to use multiple
routers per AS, and such extensions may result in significantly higher path
diversity than \cite{MuFeMaRoUh06} reported.

\section{Applying Heuristics to the Data}\label{sec:results}

\begin{figure}
\centering
\subfigure[\scriptsize Persistency distribution of paths in $\cup_{k = 1}^{15} P_k$.]{
    \label{fig_miscnf1}
    \begin{minipage}{3.5cm}
    \resizebox{4.3cm}{!}{
    \includegraphics[width=1.5in,angle=-90]{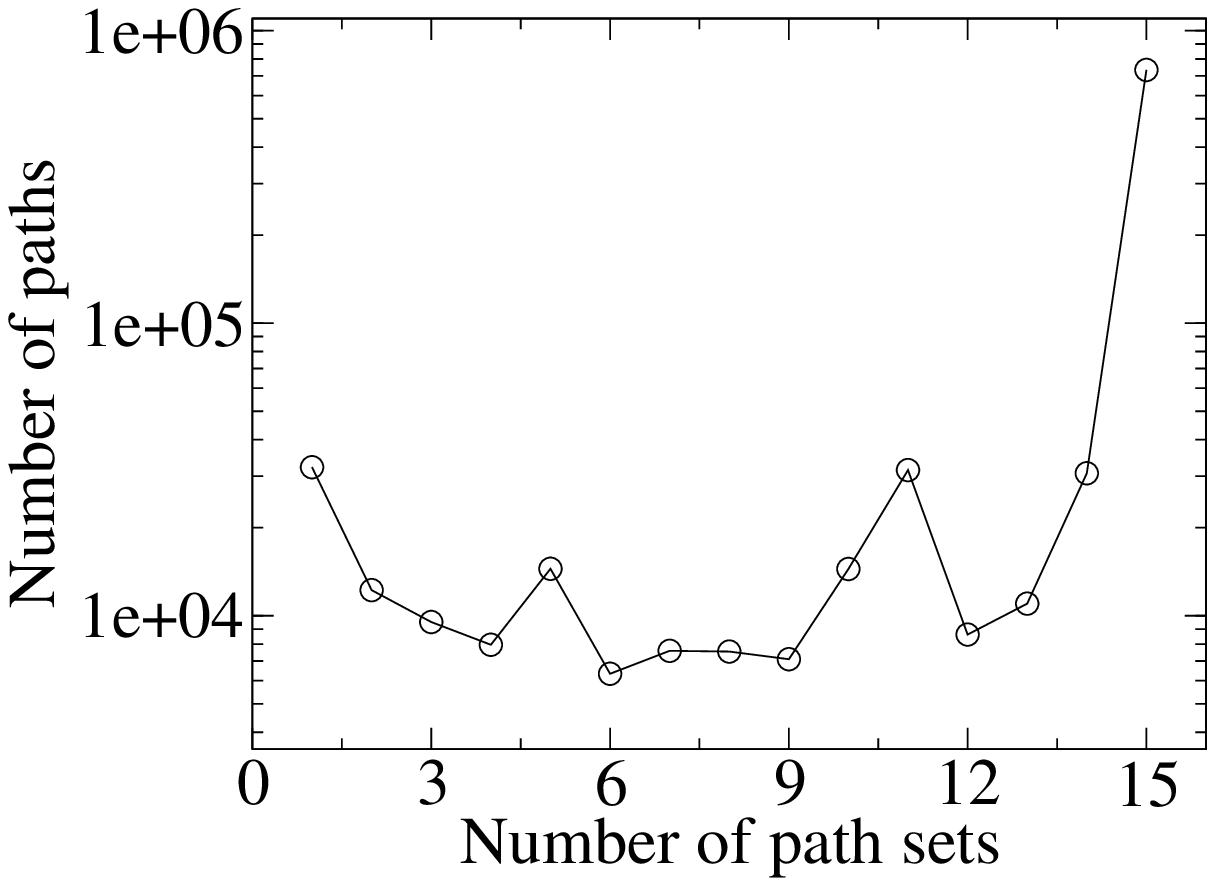}
    }
    \end{minipage}
    }
\subfigure[\scriptsize Numbers of paths, links, and ASs in the stable path set
    vs.\ the same numbers averaged over~$P_k$, $k = 1\ldots 15$.]{
    \label{fig:CompStable}
    \begin{minipage}{1.5in}
    \scriptsize
    \resizebox{3.2cm}{!}{
        \begin{tabular}{|l|c|c|}
        \hline
        & Stable & Average\\
        & paths ($P$)  & over $P_k$\\
        & (decrease) & \\
        \hline
        \hline
        \multirow{2}{*}{Paths} & \paths & \multirow{2}{*}{\pathsav }\\
        & (-\pathsred \%) & \\
        \hline
        \multirow{2}{*}{Links} & \linksnum & \multirow{2}{*}{\linksav } \\
        & (-\linksnumred \%) & \\
        \hline
        \multirow{2}{*}{ASs} & \nodes & \multirow{2}{*}{\nodesav }\\
        & (-\nodesred \%) & \\
        \hline
        \end{tabular}
    }
    \end{minipage}
}
\caption{\scriptsize Stable paths set~$P$ vs.\ unprocessed path sets~$P_k$.}
\end{figure}

\subsection{Collecting and sanitizing the data}
\label{sec:results:data}

We first construct the input BGP path set~$P$ and the corresponding
graph~$G(V,E)$. We collected BGP tables from RouteViews~\cite{routeviews}, at
8-hour intervals, over the period from 03/01/2005 to 03/05/2005, for a total
of 15 BGP table instances. After cleaning out AS prepending and AS sets, each
BGP table yields a path set~$P_k$.

Invalid paths caused by BGP misconfigurations occur quite often and affect
200-1200 BGP table entries each day~\cite{sigcomm2002-misconfigs}. To mitigate
the impact of these misconfigurations, we sanitize the input data as follows.
We define the {\em persistency} of a path $p \in \cup_{k = 1}^{15} P_k$ as the
number of sets~$P_k$ containing~$p$. The persistency distribution
(Figure~\ref{fig_miscnf1}) shows that although the majority of the paths appear
in most of the 15 sets, a significant number of paths appear only in a few of
the sets. Since BGP misconfigurations are temporal events, we select as
input~$P$ to our algorithm the paths that appear in all of the 15 sets~$P_k$.
We call~$P$ the {\em stable path\/} set and the paths that are not selected
the {\em unstable path\/} set.

\begin{table}
\centering
    \scriptsize
    \caption{\scriptsize Number of unique degree-valleys and total number of degree-valleys
    found in stable and unstable path sets.}

    \label{table:valleys}
    \resizebox{8.5cm}{!}{
        \begin{tabular}{|c||c|c|c|c|}
        \hline
        \multirow{2}{*}{window $w$} & \multicolumn{2}{|c|}{Unique degree-valleys} & \multicolumn{2}{|c|}{Total degree-valleys} \\
          &  Stable paths & Unstable paths &  Stable paths & Unstable paths \\
        \hline
        \hline
        5 & 206 & 241 (+17\%) & 1290  & 2368 (+83.6\%)\\
        \hline
        10 & 167 & 208 (+24.6\%) & 1178  & 2290 (+94.4\%)\\
        \hline
        15 & 150 & 190  (+26.6\%) & 1135 & 2135 (+88.1\%)\\
        \hline
        20 & 141 & 171  (+21.2\%) & 1119 & 1609 (+43.8\%)\\
        \hline
        \end{tabular}
    }
\end{table}

In Figure~\ref{fig:CompStable} we compare the number of paths, links and ASs in
the stable paths set with the corresponding averages over~$P_k$. Even though
$P$ is \pathsred \% smaller than the average size of~$P_k$, our filtering of
unstable paths does not entail significant information loss in terms of number
of links (\linksnumred \% reduction) and ASs (\nodesred \% reduction).

We also verify that the unstable paths include more
non-hierarchical degree sequences, which is often an indication of a
misconfiguration~\cite{sigcomm2002-misconfigs}, than the stable paths.
We call a {\em degree-valley\/} any AS sequence A-B-C
with degrees~$d_A$, $d_B$, and~$d_C$, such that both~$d_A$ and~$d_C$ are larger
than~$d_B$ plus a small margin constant~$w$: $d_A,d_C > d_B + w$. (The small
margin~$w$ is added to filter out trivial differences between $d_B$ and $d_A$,
$d_C$.) Then, for both the stable and unstable path sets, we 100 times randomly
select 10,000 paths and count the number of degree-valleys for different~$w$.
Table~\ref{table:valleys} shows the average number of unique degree-valleys
and the average of the total number of degree-valleys in the selected paths.
The number of degree-valleys in the unstable paths is between~17\% and~94.4\%
larger than in the stable paths.

\subsection{Inferring AS relationships}
\label{sec:results:relat}

{\bf {\em s2s relationships.}}
To infer s2s relationships in our graph we use the RIPE, ARIN, and APNIC
databases, collected on 06/10/2004.\footnote{Since we extract from these
databases the information that changes slowly with time, the date of the
database dump is not critically important.}
We analyze the databases according to the methodology outlined in
section~\ref{sec:algos:s2s} and find 1,943 organizations that own multiple AS
numbers. We then examine the input graph~$G$ and discover \linksss\ edges
between ASs that belong to the same organization ($|S|$ = \linksss ).

\vskip 6pt
{\bf {\em c2p relationships.}}
We remove edges inferred as s2s from~$E$ and apply our methodology detailed in
section~\ref{sec:algos:c2p} to the remaining links $E \setminus S$. Our
implementation uses parts of the code from EHS~\cite{ErHaSch02}, the LEDA~v4.5
software library~\cite{leda}, and a publicly available SDP solver
DSDP~v4.7~\cite{dsdp}. We compute orientations of the edges in~$E \setminus S$
for different values of~$\alpha$, sampling densely the interval between 0 and 1.
Recall that when~\mbox{$\alpha=1$}, our problem formulation is equivalent to
the original ToR formulation, whereas \mbox{$\alpha=0$}
corresponds to entirely degree-based relationship inference.

To evaluate the computed orientations, we introduce a metric called
{\em reachability}. We define reachability of an AS~$X$ as the number of ASs one
can reach from this AS traversing only p2c edges. The reachability of an AS has
the following two properties: 1)~it is determined entirely from the inferred c2p
relationships; and 2)~it induces a natural hierarchy of ASs based on the size of
their customer trees. These two properties enable us to perform an initial
validation of the inferred c2p relationships by matching the top ASs in the
calculated hierarchy against the empirically known largest ISPs in the Internet.

We sort all ASs by their reachability, and group ASs with the same reachability
into {\em levels}. ASs at the highest level have the largest trees of customer
ASs. ASs at the lowest level have the smallest reachability. We then define the
position {\em depth\/} of an AS~$X$ as the number of ASs at the reachability
levels above the level of the AS~$X$.  We define the position {\em width\/} of
an AS~$X$ as the number of ASs at the same level as the AS~X.

\begin{table*}
    \centering
    \caption{\scriptsize
    The reachability-based hierarchy of ASs and percentage of invalid
    paths as functions of~$\alpha$. For different values of~$\alpha$,
    we show the position {\em depth\/} (the number of AS at the
    levels above) and {\em width\/} (the number of ASs at the same level)
    for the ten ASs that occupy the top five positions when~$\alpha$ takes
    its two extreme values: \mbox{$\alpha = 0$} and \mbox{$\alpha = 1$}.
    The AS numbers are matched to AS names using the WHOIS databases.}
    \label{table:ranking}
\resizebox{18cm}{!}{
\begin{tabular}{|c|c|c|c|cc|cc|cc|cc|cc|cc|}
\hline
\multicolumn{4}{|c|}{ }
&\multicolumn{2}{c|}{ $\alpha = 0.00 $} &\multicolumn{2}{c|}{ $\alpha = 0.01 $} &\multicolumn{2}{c|}{ $\alpha = 0.05 $} &\multicolumn{2}{c|}{ $\alpha = 0.10 $} &\multicolumn{2}{c|}{ $\alpha = 0.50 $} &\multicolumn{2}{c|}{ $\alpha = 1.00 $} \\ \hline
\multicolumn{16}{|c|}{ Percentage of invalid paths }
\\ \hline
\multicolumn{4}{|c|}{  }
&\multicolumn{2}{c|}{  12.75\% } &\multicolumn{2}{c|}{  1.79\% } &\multicolumn{2}{c|}{  0.69\% } &\multicolumn{2}{c|}{  0.46\% } &\multicolumn{2}{c|}{  0.36\% } &\multicolumn{2}{c|}{  0.33\% } \\ \hline
\multicolumn{16}{|c|}{ Top of reachability based hierarchy }
\\ \hline
 & AS \# &  name & degree & dep. & wid. & dep. & wid. & dep. & wid. & dep. & wid. & dep. & wid. & dep. & wid. \\ \hline
\hline
 \multirow{5}{*}{\rotatebox{90}{ $\alpha = 0$}}  & 701 & UUNET & 2334  & 0&1  & 1&1  & 0&105  & 0&120  & 2&201  & 11&319 \\ \cline{2-16} 
& 7018 & AT\&T & 1911  & 1&1  & 2&1  & 0&105  & 0&120  & 2&201  & 11&319 \\ \cline{2-16} 
& 1239 & Sprint & 1703  & 2&1  & 0&1  & 0&105  & 0&120  & 2&201  & 11&319 \\ \cline{2-16} 
& 3356 & Level 3 & 1228  & 3&1  & 3&1  & 0&105  & 0&120  & 2&201  & 11&319 \\ \cline{2-16} 
& 209 & Qwest & 1105  & 4&1  & 4&1  & 0&105  & 0&120  & 2&201  & 11&319 \\ \hline 
\hline
 \multirow{5}{*}{\rotatebox{90}{ $\alpha = 1$}}  & 14551 & UUNET & 35  & 128&1  & 137&2  & 138&1  & 151&1  & 260&2  & 0&1 \\ \cline{2-16} 
& 13987 & IBASIS Inc. & 3  & 1792&955  & 1802&963  & 1830&976  & 1847&971  & 1885&966  & 1&2 \\ \cline{2-16} 
& 8631 & Routing Arbiter & 48  & 108&1  & 123&1  & 122&2  & 0&120  & 0&1  & 1&2 \\ \cline{2-16} 
& 23649 & Hong Kong Teleport & 4  & 1792&955  & 1802&963  & 899&121  & 916&121  & 967&119  & 3&8 \\ \cline{2-16} 
& 4474 & Village Communications & 2  & 2747&16136  & 2765&16118  & 2806&16077  & 2818&16065  & 2&201  & 3&8 \\ \hline 
\end{tabular}

}
\end{table*}

In Table~\ref{table:ranking} we examine the top five ASs in the hierarchies
calculated for the two extreme cases, \mbox{$\alpha=0$} and \mbox{$\alpha=1$}.
When \mbox{$\alpha=0$}, the well-known ISPs: UUNET, AT\&T, Sprint, Level~3, and
Qwest occupy the top five positions in the hierarchy. On the other hand, when
\mbox{$\alpha=1$}, these positions are taken by ASs of very small degrees,
e.g., AS13987 of degree 3. The columns in the table track the position of these
ASs in the hierarchies induced for~$\alpha$ equal to 0, 0.01, 0.05, 0.1, 0.5,
and 1. We observe that as $\alpha$ gets closer to~1, the well-known ASs drift
away from the top of the hierarchies, thus highlighting an increasingly stronger
deviation from reality. This deviation is maximized when~\mbox{$\alpha=1$}
(the original ToR formulation), demonstrating the limitations of AS relationship
inference based solely on maximization of the number of valid paths.

The induced hierarchies suggest that solutions with values of~$\alpha$ close
to~1 are incorrect since they propel small ASs to the top of the hierarchy,
while well-known ISPs sink to lower positions. On the other hand, the percentage
of invalid paths, listed at the top of Table~\ref{table:ranking}, attains its
maximum of~$12.75$\% when \mbox{$\alpha = 0$}. The latter observation suggests
that the solution with \mbox{$\alpha = 0$} is also incorrect since a large number of paths
violates the valley-free routing model. Taken together, these two observations
indicate that intermediate values of~\mbox{$\alpha$} yield best solutions to
our multiobjective optimization resulting both in realistic hierarchies and in
small numbers of invalid paths. We emphasize that there is no oracle, intrinsic
to the multiobjective optimization problem formulation, that would reveal
the proper balance between the two objectives and the corresponding ``right''
value of~$\alpha$. As is typically the case with multiobjective optimization~\cite{CoSi03}, we
must exercise our external expert knowledge of data specifics to sift out
the most realistic relative weight of the objectives. For our problem, we
formalize this expert insight as follows: we search for the value of~$\alpha$
corresponding to the smallest percentage of invalid paths among all the
solutions that have only well-known ISPs at the top of the hierarchy. In our
experiments, this most realistic value of~$\alpha$ is~$0.01$
(cf.~Table~\ref{table:ranking}).

\vskip 6pt
{\bf {\em p2p relationships.}}
We implement our p2p heuristic detailed in section~\ref{sec:algos:p2p},
using the QUALEX~\cite{qualex} solver to approximate the MWIS problem.
We then infer p2p relationships in the AS topology~$G$ and construct the set~$F$
of p2p links. After removing from~$F$ the set~$S$ of s2s links, we obtain our
final answer that contains~$\linkspp$ p2p links ($|F \setminus S| = \linkspp$).

\vskip 6pt
Table~\ref{table:sum} summarizes our results for the whole graph $G(V,E)$.

\begin{table}
\centering
   \scriptsize
    \caption{\scriptsize Summary statistics of the inferred relationships.}

    \label{table:sum}

        \begin{tabular}{|c||c|c|c|c|}
        \hline
        & Total & c2p links & p2p links & s2s links \\
        & $|E|$ & $|E \setminus F \setminus S|$ & $|F \setminus S|$ & $|S|$ \\
        \hline
        \hline
        number of links & $\linksnum\ $ & $\linkscp\ $ & $\linkspp\ $ & $\linksss\ $ \\
        \hline
        percentage & $100$\% & $\linkscpper $\% & $\linksppper $\% & $0.46$\% \\
        \hline
        \end{tabular}
\end{table}

\subsection{Repository of AS Relationships and AS Rank}
\label{sec:rank}

To make our results easily accessible and practically useful for the community,
we automated our inference heuristics. We archive the inferred AS relationships
on a weekly basis and make them available for download at the AS relationship
data repository~\cite{as-rel-data}.

We also created an interactive web site~\cite{as-ranking} where we apply our
automated relationship inferences to rank ASs based on their {\em customer cones}.
We define the customer cone of an AS A as the AS A itself plus all the ASs that
it can reach ``for free'', that is, following only p2c and s2s links.
In other words, AS A's customer cone is A, plus A's customers,
plus its customers' customers, and so on. We use the following three metrics
to measure the size of customer cones: the number of ASs in the cone, the
number of unique prefixes advertised by these ASs, and the number of /24 blocks
in the union of these prefixes.

AS ranking is valuable not only for conceptual understanding of relative
importance of Internet players, but also for network vendors and operators
in prioritizing their customer lists and in solving other practical tasks.
Users of our AS ranking have an option to group multiple sibling
ASs into one organizational entity by specifying sibling groups either
from the IRRs data, or as user-provided sibling lists.

\section{Survey and Validation}
\label{sec:sur}

Measuring, understanding, and modeling AS relationships in the Internet
are challenging tasks hampered by the fact that these
relationships are sensitive business information and generally considered
private by ISPs.  Nevertheless, without validation against truth,
we have no way of evaluating the integrity of our heuristics.

Most of the previous works relied on implicit validation. However, indirect
approaches are not always reliable. For example, the authors
of~\cite{SuAgReKa02,DiBaPaPi03, ErHaSch02,MaQiEaZh05} used the number
of valid paths as an indicator of the accuracy of the inferred relationships.
As we discussed in section~\ref{sec:algos:pre}, a large number of valid paths
does not necessarily result in a large number of correctly inferred AS
relationships.

In contrast with previous works, we augment our validation based on
implicit metrics (e.g., reachability, section~\ref{sec:results:relat}) with
the explicit data that we collected via private communications with engineers
from the ASs under observation.

We contacted several ASs ranging from large continental or national ISPs,
to content providers, and university networks. We sent
the list of AS relationships that we inferred for a given AS to this AS's
network administrator, peering negotiator, informed engineer, or researcher.
We included three questions in our email inquiry:\footnote{
We also offered to sign a non-disclosure agreement (NDA) that protected
peering information from being released to the public and regulated
our data analysis to anonymizing the participating ASs.
Only one organization (a government agency) required an NDA and two
commercial ISPs did not have a policy in place (or the policy was not)
to deal with such requests. They still helpfully provided us with general
answers regarding what percent of peers we inferred incorrectly.}

\begin{description*}
\item[Q1:] For the listed inferred AS relationships, specify how many are incorrect,
and what are the correct types of the relationships that we mis-inferred?

\item[Q2:] What fraction of the total number of your AS neighbors is included in our list?

\item[Q3:] Can you describe any AS relationships, more complex than c2p, p2p, or s2s,
that are used in your networks?
\end{description*}

We performed the survey in the period between 06/07/05 and 06/30/05 and received
answers from \assnum\ out of the~78 ASs we contacted. Among these, 5 were tier-1
ISPs, 13 were smaller ISPs, 19 were universities, and 1 was a content provider.
These ASs reported to us the true relationship types for~\seentruenum\ of our
inferred AS relationships. The universities reported only~54 of those~\seentruenum\
relationships, whereas all the remaining relationships came from the ISPs and
the content provider.
The BGP-derived AS degrees for the universities ranged from~1 to~8,
while for the remaining ASs it ranged from~1 to almost~2000.

\subsection{Validation of inferred AS relationships}
\label{sec:sur:val}

\begin{table}
\centering
    \caption{\scriptsize Validation of the inference results using the survey data.
    Each row shows the total number, number of correct, and percentage of correct
    inferred AS relationships.}

    \resizebox{8.5cm}{!}{

    \begin{tabular}{|c||c||c|c|c|}
    \hline
            & \multirow{2}{*}{links} & inferred & inferred & inferred\\
            &  & c2p links& p2p links& s2s links\\
    \hline
    \hline
    total number of      & \seentruenum &
            \seeninfcppc &
            \seeninfpp &
            \seeninfss \\
    \hline
    number of correct  & \validnumtot &
            \validnumcppc &
            \validnumpp &
            \validnumss    \\

    \hline
    percentage of correct  & \validinfpertot &
            \validinfpercppc &
            \validinfperpp &
            \validinfperss \\
    \hline
    \end{tabular}
    }
    \label{table:val}
\end{table}

We validate our heuristics by counting the number of correctly inferred AS
relationships. Among the \seentruenum\ verified AS relationships,
\seentruecppcper\ were c2p, \seentrueppper\ were p2p, and \seentruessper\
were s2s. Table~\ref{table:val} demonstrates
that our heuristics correctly infer \validinfpercppc\ of c2p, \validinfperpp\ of
p2p, and \validinfperss\ of s2s relationships. The total percentage of correctly
inferred AS relationships is \validinfpertot. This accuracy level demonstrates
that our heuristics produce reliable and veracious inferences of the true types
of AS relationships in the Internet.

The data in our survey bear certain limitations and our results should be
interpreted accordingly. First, the self-selection aspect of the sampling of
ASs may induce biases into the resulting statistics. Second, the obtained
\seentruenum\ links with confirmed relationships represent~9.7\% of
the total number of links in our data. While acknowledging these limitations,
we note that providing rigorous validation of inferred AS relationships is an
extremely challenging task because of the difficulty in collecting ground-truth
data against which one can check the inferences.

\subsection{Missing AS links}

In this section we analyze the relationships of the full set
of adjacencies of the participating ASs, including the links that we do not
see in BGP tables and, consequently, in our graph. The second question in our
survey asks ASs for the ratio of the number of their neighbors in our AS
topology data to the total number of AS neighbors they actually have. Out of
the \assnum\ ASs, \asswfulldata\ (3 of which were tier-1 ISPs) provided us not
only with this ratio, but also with the types of relationships their ASs have
with the missing neighbors. Out of the total of \truelinks\ true reported
adjacencies, the BGP tables observe only~\seenlinks. This finding agrees with
the conclusion of previous works~\cite{ChaGoJaSheWi04,RaCo06} that a significant
number of existing AS connections remain hidden from most BGP routing tables.

\begin{figure}
    \centering
    \includegraphics[width=1.9in]{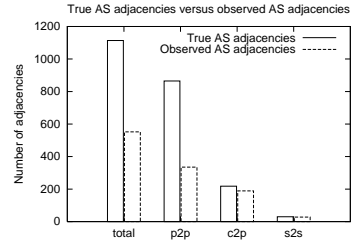}
    \caption{\scriptsize Numbers of true and observed AS links for
     different types of AS relationships in the survey.}
    \label{fig:val1}
\end{figure}

To improve our understanding of the missing AS links, we analyze the true
relationships of these links.
Figure~\ref{fig:val1} illustrates the per-relationship breakdown of the true
and observed adjacencies of the ASs.
It shows that we only see \trueseenppdiffper\ out of the \truepplinks\ p2p
links, whereas we see \trueseencppcdiffper\ out of the \truecppclinks\ c2p
links, and \trueseenssdiffper\ of the \truesslinks\ s2s links. This gap
demonstrates that BGP-derived AS topologies miss predominantly p2p links.

The reasons for this bias stem from the intrinsic nature of p2p
relationships. In a p2p relationship, prefixes learned from a peer AS are not
advertised to any providers. Consequently, a link between two p2p ASs is not
seen (as a part of some AS path) at any upstream ASs. It follows that we
can only observe a p2p link in the BGP tables of the customers or siblings of
the p2p ASs. The periphery of the Internet has many small interconnecting
ASs. Thus, in order to observe p2p links in the periphery, we should have a significant number and variety of BGP
tables from these small ASs. BGP tables with small number of data feeds alone do
not provide representative statistics of p2p links.

Figure~\ref{fig:val1} also shows that the majority of the~\truelinks\ true
adjacencies are in reality p2p: \truepplinks\ (\ppper) are p2p, while only
\truecppclinks\ (\cppcper) and \truesslinks\ (\ssper) are c2p and s2s,
respectively. We thus face a large number of p2p relationships which appear to
be very popular among small and medium size ASs. Interestingly,
some tier-1 ISPs have several dozens or even hundreds of p2p relationships,
frequently with ASs of smaller size.

\begin{figure}
    \begin{center}
    \includegraphics[width=2.3in]{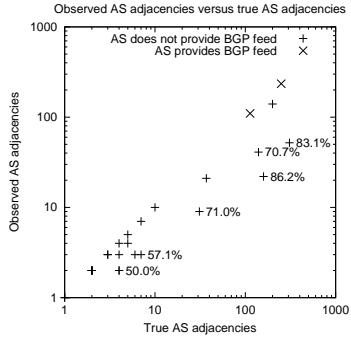}
    \caption{\scriptsize True vs.\ observed degrees of the surveyed ASs.
    We mark 1) the percentage of missed links for a few ASs with the highest
    values of this percentage; and 2) the ASs that did (not) provide feeds
    in our BGP data.}
    \label{fig:val2}
    \end{center}
\end{figure}

Next, we seek to evaluate how representative the BGP-derived AS degrees
are of the true AS degrees. In Figure~\ref{fig:val2} we plot the number
of true AS adjacencies of the surveyed ASs versus the number of AS
adjacencies derived from our BGP data.
At the bottom-left corner of the diagram, 20 ASs\footnote{
Note that points (1,1) and (2,2) in the figure correspond to more than one AS.}
that are mainly university networks, have their true numbers of adjacencies
close or identical to the measured numbers of adjacencies. We find that most
of the adjacencies of these small ASs are c2p links. As we have seen above,
our AS topology captures c2p links relatively well. Examining the rest of the
diagram, we first observe that the percentage of missed adjacencies can be as
large as~86.2\%. The degrees for most of the highly connected ASs are
under-sampled, half of them missing more than~70.5\% links. Further examination
of the missed AS links reveals that most of them are of p2p type, which is
consistent with Figure~\ref{fig:val1}.

Our results confirm the common assumption that p2p relationships, while widespread in the Internet,
are not amenable to observation from few BGP feeds and can render BGP-derived AS
degrees significantly smaller than the true AS degrees. On one hand, the identified deficiencies
should inspire further pursuit of representative statistics on the number of
p2p links, for example via the deployment of distributed measurement
infrastructures~\cite{dimes-ccr}. On the other hand, we emphasize that
these missing links do not qualitatively change a set of frequently-used
{\em statistical\/} characteristics of the BGP-derived AS
topologies~\cite{DaAlHaBaVaVe05,MaKrFoHuDiclVa06,CoGoWo06}.

\subsection{Complex AS relationships}
\label{sec:sur:ana}

The last question of our survey asks about more complex configuration scenarios
the AS may be using. From the responses we learn that although the majority of
AS relationships are simply c2p or p2p, in few cases their configurations are
either more specialized versions of the basic c2p or p2p types, or a hybrid of
c2p and p2p (c2p/p2p).

For example, the backup provider relationship is a specialized variant of the
basic c2p relationship. In this case, a customer AS has a c2p relationship with
a provider AS, but this relationship only allows traffic to flow
during an emergency situation such as a disruption of connectivity to the
main upstream provider of the customer AS. Hence, the backup provider
relationship is a temporally conditioned version of the c2p relationship.

A hybrid c2p/p2p relationship occurs when two ISPs interconnect at multiple
peering points and have different types of relationships at these points. For
instance, two ISPs can have a p2p relationship at a peering point in Europe and
a c2p relationship at a peering point in the U.S. Another flavor of a hybrid
c2p/p2p relationship is when two ASs have different types of relationships for
different IP prefixes. In this case the ISPs may have a p2p relationship for
one set of IP prefixes and a c2p relationship for another set of IP prefixes.
These examples of hybrid c2p/p2p relationships illustrate that AS relationships
may involve also spatial and/or prefix-based aspects.

In other words, based on the configuration descriptions we collected
in our survey, we conclude that AS relationships vary across the
following three dimensions: space, time, and prefix.  Therefore, to
fully characterize a relationship between a pair of ASs, including
more complex relationship scenarios, one has to gain access to
information identifying the ASs' policy configurations per peering
location(s), per time, and per prefix.  Although limited per-prefix and
per-time data are presently available, identifying more complex
relationships for the complete Internet AS topology is a formidable task as it
likely requires significantly more sources of
more detailed data than currently available.

A natural question that arises is how a c2p, p2p, or s2s inference for
an AS link, which in fact is a more complex relationship, distorts reality and how
prominent this artifact is. For a backup relationship, a c2p, p2p, or
s2s inference misses the temporal component of the relationship.  For
a hybrid c2p/p2p relationship, a c2p or p2p inference misses one part
of the hybrid relationship. Such artifacts, however, do not occur
often. Indeed, more complex relationships are likely to exist only between
large ASs. However, most AS links in the Internet connect small ASs to large
ASs or connect small ASs to each other~\cite{MaKrFoHuDiclVa06}. Such AS pairs are
known to employ consistent routing policies over their usually single or sometimes
multiple peering points.

\section{Conclusion}
\label{sec:con}

The relationships among ASs in the Internet represent the outcome of policy
decisions governed by technical and business factors of the global Internet
economy. Precise knowledge of these relationships is therefore an essential
building block needed for any reliable and effective analysis of technical and
economic aspects of the global Internet, its structure, and its growth.

In this work we introduced novel heuristics that significantly improve the
state-of-the-art in inferring c2p relationships and carefully address the
particularly difficult problems of inferring p2p and s2s relationships from
currently available data.

In comparison with previous studies that primarily used implicit validation of
inferred AS relationships, we go a step further. In addition to implicit validation,
we make an effort to collect explicit ground-truth data via direct
communication with ASs. Using the true relationships of \seentruenum~links
we confirmed that our heuristics achieve very high accuracy of \validinfpercppc~(c2p),
\validinfperpp~(p2p), and \validinfperss~(s2s) of correctly inferred
relationships, with the overall accuracy being~\validinfpertot.
Given the overall difficulty of validating inference results and that
surveys like the one in this paper tend to be extremely involved
procedures in practice, we hope that our work will serve to cast
ponderable confidence on such inference studies.

Using the data of our survey we followed previous studies~\cite{ChaGoJaSheWi04,RaCo06}
in finding that measured AS topologies miss a significant number of AS links.
We take this result further by verifying the commonly held assumption
that most of the missing links are of p2p type.

Easy access to accurate AS relationship data is essential to a variety
of studies dealing with aspects of Internet architecture and policy.
To support the research community with as objective data as possible,
we have automated our heuristics and calculate and archive AS
relationships on a weekly basis~\cite{as-rel-data}. As an example of
using the inferred relationships we provide a ranking of ASs~\cite{as-ranking}.

From the perspective of empirical research, the global Internet compares to an
economy or an ecosystem. As such, cross-disciplinary approaches that combine
knowledge of the Internet macroscopic structure with insights into its
economics and policy are required to advance our understanding of its technical
and economical viability.
We believe our work significantly benefits Internet research that
strives to build more encompassing models validated against reliable and
accurate data.

\section*{Acknowledgment}
This work was supported by CISCO Systems, Inc., and by the NSF grants
CNS-0434996, CNS-0427700, ANI-0240477, SCI-0427144, and ECS-0225417.

\scriptsize
\bibliographystyle{abbrv}
\bibliography{bib}

\begin{thebibliography}{10}

\bibitem{leda}
{Algorithmic Solutions Software GmbH}.
\newblock {L E D A} library, 2004.
\newblock \url{http://www.algorithmic-solutions.com/enleda.htm}.

\bibitem{AlDaBaVe05}
I.~Alvarez-Hamelin, L.~Dall'Asta, A.~Barrat, and A.~Vespignani.
\newblock $k$-core decomposition: A tool for the analysis of large scale
  {Internet} graphs.
\newblock \url{arXiv:cs.NI/0511007}.

\bibitem{AlDaBaVe04}
I.~Alvarez-Hamelin, L.~Dall'Asta, A.~Barrat, and A.~Vespignani.
\newblock $k$-core decomposition: A tool for the visualization of large scale
  networks.
\newblock \url{arXiv:cs.NI/0504107}.

\bibitem{DiBaPaPi03}
G.~D. Battista, M.~Patrignani, and M.~Pizzonia.
\newblock Computing the types of the relationships between {A}utonomous
  {S}ystems.
\newblock In {\em IEEE INFOCOM}, 2003.

\bibitem{dsdp}
S.~Benson, Y.~Ye, and X.~Zhang.
\newblock A dual-scaling algorithm for semidefinite programming, 2004.
\newblock \url{http://www-unix.mcs.anl.gov/~benson/dsdp/}.

\bibitem{qualex}
S.~Busygin.
\newblock {QU}ick {AL}most {EX}act maximum weight clique/independent set
  solver.
\newblock \url{http://www.busygin.dp.ua/npc.html}.

\bibitem{as-rel-data}
{CAIDA}.
\newblock {AS Relationships Data}.
\newblock Research Project.
\newblock \url{http://www.caida.org/data/active/as-relationships/}.

\bibitem{as-ranking}
{CAIDA}.
\newblock Automated {Autonomous System} ({AS}) ranking.
\newblock Research Project.
\newblock \url{http://as-rank.caida.org}.

\bibitem{ChaGoJaSheWi04}
H.~Chang, R.~Govindan, S.~Jamin, S.~J. Shenker, and W.~Willinger.
\newblock Towards capturing representative {AS}-level {I}nternet topologies.
\newblock {\em Computer Networks Journal}, 44:737--755, April 2004.

\bibitem{CoGoWo06}
R.~Cohen, M.~Gonen, and A.~Wool.
\newblock Bounding the bias of tree-like sampling in {IP} topologies.
\newblock \url{arXiv:cs.NI/0611157}.

\bibitem{CoSi03}
Y.~Collette and P.~Siarry.
\newblock {\em Multiobjective Optimization: Principles and Case Studies}.
\newblock Springer-Verlag, Berlin, 2003.

\bibitem{DaAlHaBaVaVe05}
L.~Dall'Asta, I.~Alvarez-Hamelin, A.~Barrat, A.~V\'{a}zquez, and A.~Vespignani.
\newblock Exploring networks with traceroute-like probes: Theory and
  simulations.
\newblock {\em Theoretical Computer Science, Special Issue on Complex
  Networks}, 2005.

\bibitem{DiKrHuClRi05}
X.~Dimitropoulos, D.~Krioukov, B.~Huffaker, kc~claffy, and G.~Riley.
\newblock Inferring {AS} relationships: Dead end or lively beginning?
\newblock In {\em Proceedings of 4th Workshop on Efficient and Experimental
  Algorithms (WEA' 05)}, May 2005.

\bibitem{ErHaSch02}
T.~Erlebach, A.~Hall, and T.~Schank.
\newblock Classifying customer-provider relationships in the {I}nternet.
\newblock In {\em Proceedings of the {IASTED} International Conference on
  Communications and Computer Networks ({CCN})}, 2002.

\bibitem{Gao01}
L.~Gao.
\newblock On inferring {Autonomous System} relationships in the {Internet}.
\newblock In {\em {IEEE/ACM} Transactions on Networking}, December 2001.

\bibitem{GaJo79}
M.~R. Garey and D.~S. Johnson.
\newblock {\em Computers and Intractability: A Guide to the Theory of
  {NP}-Completeness}.
\newblock W. H. Freeman and Company, San Francisco, 1979.

\bibitem{GoWi95}
M.~X. Goemans and D.~P. Williamson.
\newblock Improved approximation algorithms for maximum cut and satisfiability
  problems using semidefinite programming.
\newblock {\em Journal of the {ACM}}, 42(6):1115--1145, 1995.

\bibitem{MaKrFoHuDiclVa06}
P.~Mahadevan, D.~Krioukov, M.~Fomenkov, B.~Huffaker, X.~Dimitropoulos,
  k.~claffy, and A.~Vahdat.
\newblock The {Internet} {AS}-level topology: Three data sources and one
  definitive metric.
\newblock {\em {ACM} Computer Communications Review}, 36(1):17--26, 2006.

\bibitem{sigcomm2002-misconfigs}
R.~Mahajan, D.~Wetherall, and T.~Anderson.
\newblock Understanding {BGP} misconfiguration.
\newblock In {\em ACM SIGCOMM}, August 2002.

\bibitem{MaQiEaZh05}
Z.~M. Mao, L.~Qiu, J.~Wang, and Y.~Zhang.
\newblock On {AS}-level path inference.
\newblock In {\em SIGMETRICS}, 2005.

\bibitem{routeviews}
D.~Meyer.
\newblock University of {O}regon {R}oute {V}iews {P}roject, 2004.

\bibitem{MuFeMaRoUh06}
W.~Muhlbauer, A.~Feldmann, O.~Maennel, M.~Roughan, and S.~Uhlig.
\newblock Building an {AS}-topology model.
\newblock In {\em ACM SIGCOMM}, 2006.

\bibitem{RaCo06}
D.~Raz and R.~Cohen.
\newblock The {I}nternet dark matter: on the missing links in the {AS}
  connectivity map.
\newblock In {\em INFOCOM}, 2006.

\bibitem{rimondini02}
M.~Rimondini.
\newblock Statistics and comparisons about two solutions for computing the
  types of relationships between {A}utonomous {S}ystems, 2002.
\newblock
  \url{http://www.dia.uniroma3.it/~compunet/files/ToR-solutions-comparison.pdf%
}.

\bibitem{dimes-ccr}
Y.~Shavitt and E.~Shir.
\newblock {DIMES}: Let the {Internet} measure itself.
\newblock {\em Computer Communication Review}, 35(5), 2005.

\bibitem{SuAgReKa02}
L.~Subramanian, S.~Agarwal, J.~Rexford, and R.~H. Katz.
\newblock Characterizing the {Internet} hierarchy from multiple vantage points.
\newblock In {\em IEEE INFOCOM}, 2002.

\bibitem{SuKaEe05}
L.~Subramanian, M.~Caesar, C.~T. Ee, M.~Handley, M.~Mao, S.~Shenker, and
  I.~Stoica.
\newblock {HLP}: A next generation inter-domain routing protocol.
\newblock In {\em ACM SIGCOMM}, 2005.

\bibitem{TaDoGoJaWiSh01}
H.~Tangmunarunkit, J.~Doyle, R.~Govindan, S.~Jamin, W.~Willinger, and
  S.~Shenker.
\newblock Does {AS} size determine {AS} degree?
\newblock {\em {ACM} Computer Communication Review}, October 2001.

\bibitem{UhMaMu06}
S.~Uhlig, O.~Maennel, and W.~Muelbauer.
\newblock Modeling as a necessary step for understanding {I}nternet-wide route
  propagation.
\newblock In {\em {WIRED}: Workshop on {I}nternet Routing Evolution and
  Design}, 2006.

\bibitem{XiaGao04}
J.~Xia and L.~Gao.
\newblock On the evaluation of {AS} relationship inferences.
\newblock In {\em IEEE GLOBECOM}, 2004.

\end{thebibliography}

\end{document}